%% file: main.tex
\documentclass[twocolumn,trackchanges]{aastex63}

\usepackage{graphicx}
\usepackage{url}
\usepackage{hyperref}

\received{...}
\revised{...}
\accepted{...}
\submitjournal{AJ}

\begin{document}

\title{Spatially resolved velocity structure in jets of DF Tau and UY Aur A}

\correspondingauthor{Anastasiia V. Uvarova}
\email{avu@mit.edu}

\author{Anastasiia V Uvarova}
\affil{MIT, Kavli Institute for Astrophysics and Space Research, 77 Massachusetts Avenue, Cambridge, MA 02139, USA}

\author[0000-0003-4243-2840]{Hans Moritz G\"unther}
\affil{MIT, Kavli Institute for Astrophysics and Space Research, 77 Massachusetts Avenue, Cambridge, MA 02139, USA}

\author{David A. Principe}
\affil{MIT, Kavli Institute for Astrophysics and Space Research, 77 Massachusetts Avenue, Cambridge, MA 02139, USA}

\author{P. Christian Schneider}
\affil{Hamburger Sternwarte, Universit\"at Hamburg, Gojenbergsweg 112, 21029, Hamburg, Germany}

\begin{abstract}
Young stars accrete mass and angular momentum from their circumstellar
disks. Some of them also drive outflows, which can be distinguished in
optical forbidden emission lines (FELs). We analyze a sample of binary
T Tauri stars observed with long-slit spectroscopy by the Hubble Space
Telescope (HST), searching for spatially resolved outflows. We detect resolved [O~{\sc i}] emission in two cases out of twenty one. In DF Tau we resolve high and medium velocity outflows in a jet and counterjet out to 60~au. The outflows are accelerated within the inner 12~au and retain a constant speed thereafter. In UY~Aur, we detect a blue- and a red-shifted outflow from UY~Aur~A, as well as a blue-shifted jet from UY~Aur~B. All of these features have been seen in [Fe~{\sc ii}] with data taken ten years apart indicating that the underlying outflow pattern is stable on these timescales.
\end{abstract}%

\section{Introduction}
\label{sect:intro}
Star formation occurs when large clouds of gas and dust collapse due to
gravity. The clouds are inhomogeneous in density and they fragment into
smaller structures, where the center of each collapsing sub-cloud may
become a star or system of stars. Very few of the resulting stars are
massive and hot. By far the largest number will evolve into late-type
stars with spectral types in the M-F range. Most stars are members of a binary or multiple multiple systems \citep[see e.g.\ review by][]{2007prpl.conf..379D}.
The infalling envelope
flattens to a circumstellar disk, making the stars visible in the
optical. Low-mass stars in this stage are called classical T Tauri stars
(CTTS). For a few Myrs, planet formation can take place before the disk
disperses. For binaries or higher-order multiple systems, the disk can
belong to an individual star or surround a close binary pair depending
on the mass and separation of the components.

Mass is accreted through these disks onto the stars. However,
conservation of angular momentum demands that some mass is ejected and
carries away the angular momentum accreted through
the disk. Mass
loss occurs through wide-angle disk winds, but in some systems we 
also see highly collimated jets \citep[see][for a
    review]{2014prpl.conf..451F}.
Based on different theoretical ideas, models of stellar winds
\citep{1988ApJ...332L..41K,2005ApJ...632L.135M}, X-winds
\citep{1994ApJ...429..781S} and disk winds
\citep{1982MNRAS.199..883B,2005ApJ...630..945A} have been proposed that can
launch mass into an outflow. Ultimately these mass flows  must be powered from
the gravitational energy released in the accretion process. The collimation of the outflow into a jet is hard to
explain without toroidal magnetic fields generated by
magnetic field that is dragged inwards in the accretion process \citep[see
  review by][]{2014ComAC...1....3L}.
 However, we do
not know in detail how energy and momentum are converted from an inflow to an outflow \citep[e.g.][and references therein]{2010ApJ...714..989M}.
Observationaly, we see that the outflow rate is roughly one tenth of the
accretion rate \citep{1990ApJ...354..687C,2008ApJ...689.1112C}, supporting a
causal relation between outflows and energy released in the accretion process. 

Full 3D magneto-hydrodynamic simulations of jet launching are numerically
challenging and early work has been limited to lower dimensions and/or ideal MHD
\citep[e.g.][]{2002ApJ...581..988C,2015MNRAS.450..481D}. For binary stars, an
added complication is the influence that one member of the binary has on the
disk and outflow from the other star. The most important parameter to determine
the strength of any gravitational torque is the binary separation. 
Simulations
for binaries separated by no more than 15~au or so indicate that no jets are
launched for high disk-orbit inclinations. For moderate inclination, the jet
axis can precess in a cone of a few degrees opening angle
\citep{2018ApJ...861...11S}. We refer the reader to the introduction in
\citet{2018ApJ...861...11S} for a review of this topic. In contrast, in wide
binaries, simulations and observations show random alignments between outflows
from both members of the binary \citep{2016ApJ...827L..11O}.

Forbidden optical emission lines (FELs) are a
good way to find and study such jets, since the stellar photosphere and
the accretion shock are too dense to contribute significantly to the emission. 
These jets typically have an onion-like
structure with a fast component at the center surrounded by increasingly
slower and less collimated components further
out \citep{2000ApJ...537L..49B}.

If a
jet is detected in several emission lines, line ratios can be used to
calculate density and ionization fraction of emission components in the
jet with typical jet densities in the range~$10^3-10^5\;\mathrm{cm}^{-3}$
\citep[e.g.][]{1993ApJ...410L..31S,1999A&A...342..717B,2000A&A...356L..41L,2013A&A...550L...1S}. In turn
the density and the velocity give mass loss rates. Different outflow
components show different velocities. For example, in the well-studied
CTTS DG Tau~\citet{2013A&A...550L...1S} find [O~{\sc i}] in a low-velocity
component (LVC, about 60~km~s$^{-1}$) which can be detected as
close as 15~au from the star and a medium-velocity component (MVC, about
130~km~s$^{-1}$) first detected at about 50 au from the source. The MVC in DG~Tau
slows down at larger distances. 
FELs can be seen up to at least 600~au from the
star in DG Tau \citep{1993ApJ...410L..31S}. FELs have also been detected in outflows from
binaries, in some cases each star launches its own jet \citep[e.g.\ in
  T~Tau][]{1999ApJ...523..709S}, in other systems it appears that the outflow
is launched from a common circumbinary disk or at least that the individual
jets from both components of the binary merge early \citep{2010ApJ...708L...5M}. 

While a single spectrum is sufficient to
detect the presence of a FEL, spatially resolved data is necessary to study
how outflows accelerate and decelerate. In this work, we reanalyze
archival data from the Hubble Space Telescope (HST) Program ID 7310 to
search for FELs that are spatially resolved.

In section~\ref{sect:obs} we describe the observations and the data reduction. Section~\ref{sect:results} presents our immediate results. We discuss the resolved emission from DF~Tau and UY~Aur in section~\ref{sect:discussion} and end with a short summary in section~\ref{Sect:summary}.

\section{Observations and data reduction}
\label{sect:obs}

\input{table_of_observations.tex}

Hubble Space Telescope (HST) Program ID 7310 targeted binary T Tauri
stars with long-slit spectroscopy using the Space Telescope Imaging
Spectrograph (STIS). The long slit is always oriented such that both
components of the binary are observed. \citet{2003ApJ...583..334H} analyze the
spectra of both stellar components to determine stellar properties and
accretion diagnostics. In this work, we aim to spatially resolve the
emission in FELs along the slit.

Table~\ref{tab:obs} lists the observations presented in this paper. The
observations for each of the stars were taken using two different gratings:
G750L and G750M, with central wavelengths of 7751~\AA{} and 6252~\AA{} with
exposure times of 360~s and 1080~s respectively. For each target system the observations were consecutive in the same orbit. Position angles are measured East of North. \citet{2003ApJ...583..334H} lists properties (including binary separation and position angle, but also derived properties like stellar mass and accretion) in tables for all targets.

Our analysis begins from the pipeline reduced
2-dimensional~\texttt{sx2} files; these files have one spectral axis and
one spatial axis for the coordinate along the slit. For each wavelength
in~\texttt{sx2} we fit a single Gaussian to approximate the spatial flux distribution (dominated by the point-spread function (PSF) of the star), taking into account regions
flagged for data quality by the pipeline. While this does not capture
all features of the instrumental PSF, it describes the signal close to the
peak of the emission well and allows a numerically stable fit of the
position of the peak.

The measured peak position  of the Gaussian changes with
wavelength in a smooth manner, both due to minor misalignment between slit and CCD as well as due to the distortion corrections done in the data
pipeline. The scatter in the fit results can be
taken as an estimate of uncertainty of our fits.
If both components of the binary system are within a few arcseconds of
each other, a fit using two Gaussians is performed. All fits are
visually inspected. 

FELs are tracers of outflows because they can only be formed in low-density
environments. We search for changes
of the fitted position of the Gaussian, i.e.\ changes in the mean
position of the emission caused by a resolved emission component offset from
the central star around the wavelength of the FELs in table~\ref{tab:searchedlines}. Limiting our search in this way reduces the
rate of false-positives.

\begin{table}
\caption{{Lines searched for spatial extension\label{tab:searchedlines}}}
\begin{center}
\begin{tabular}{cc}
\hline\hline
line & wavelength [\AA] \\
\hline
[O I] & 6302.0\\{}
[O I] & 6365.5\\{}
[O III] & 5008.240\\{}
[O III] & 4960.295\\{}
[O III] & 4364.436\\{}
[N II] & 6549.86\\{}
[N II] & 6585.27\\{}
[S II] & 6718.29\\{}
[S II] & 6732.67\\
\hline
\end{tabular}
\end{center}
\end{table}

To investigate resolved emission in more detail we subtract the contribution of
the stellar continuum similar to the method of \citet{2013A&A...550L...1S}: For
every row of in the image (in the dispersion direction) we select a region of
interest centered on the position of the FEL (15 pixels wide for the G750M data
and 2 pixels wide in the G750L data, where the FELS are not spectrally
resolved). Assuming that the PSF is only slowly changing with wavelength and
that the stellar emission around the FELs is continuum dominated, we fit a
second degree polynomial to the data to the left and right of the FEL. The
region used for the fit is 10 pixels wide in the G750M data and 5 pixels wide
in the G750L data. We subtract the value of the polynomial for every row of
data in the FEL region. The remaining images should display signatures of jets
if there are any, and can be used to examine jet profiles and intensities. 

\section{Results}
\label{sect:results}

\subsection{Significantly extended FELs}
We detected significantly extended emission only in two out of 21 objects,
DF~Tau and UY~Aur. In both cases, the extension is seen in the [O~{\sc i}]
lines, but is not significantly detected in any other
FEL. \citet{2003ApJ...583..334H} already noted that the on-source [O~{\sc i}]
line profile shows different kinematic components for both components of the
DF~Tau binary and for the primary of UY~Aur, while all other targets have
simple [O~{\sc i}] line profiles. \citet{2003ApJ...583..334H} also discuss the
\object{FS Tau} primary where they note an [O~{\sc i}] line that appears spatially
extended. FS~Tau shows an extension in our processing, however, this is due
entirely to the signal from a single pixel, which is flagged by the data
processing pipeline as having a high dark current rate. No spatial extension is seen after removing that spurious signal.

We show position-velocity-diagrams (PVD) for DF~Tau in Fig.~\ref{fig:DFTau}
and for UY~Aur in Fig.~\ref{fig:UYAur}. In the PVDs the
origin of the spatial coordinate is set on the position of the primary star. Resolved
FEL emission is seen as a bulge in the vicinity of the rest wavelength
in the emission line. Where resolved FEL emission is seen, we quantify the peak
velocity and width of the emission by fitting Gaussians to the spectrum at each
position (row of data in the PVD). See Figure~\ref{fig:Jet_Fit} for examples of
those fits at different positions.

\begin{table}
\caption{Measured fluxes\label{tab:flux}}
\begin{center}
\begin{tabular}{cccccc}
\hline\hline
source & velocity & FWHM & 6300~\AA{} & 6363~\AA{} & $\dot M$\\ 
       & [km s$^{-1}$] & [km s$^{-1}$] & \tablenotemark{a} & \tablenotemark{a} & \tablenotemark{b}\\
\hline
DF Tau & -150 & 40 & 3 & 1 &  14\\
DF Tau & -50 & 60 & 2 & .4 & 3\\
DF Tau & +150 & 40 & 3 & 5 & 14\\
UY Aur A & -150 & 30 & 0.6 & 2 & 5\\
UY Aur A & +50 & 40 & 0.3 & 1 & 0.8\\
\end{tabular}
\end{center}
\tablenotetext{a}{in units of $10^{-15}$ erg s$^{-1}$ cm$^{-2}$}
\tablenotetext{b}{in units of $10^{-10}\;M_{\odot}$ yr$^{-1}$}
\end{table}

\subsubsection{DF Tau}
\begin{figure}
\begin{center}
\includegraphics[width=0.49\textwidth]{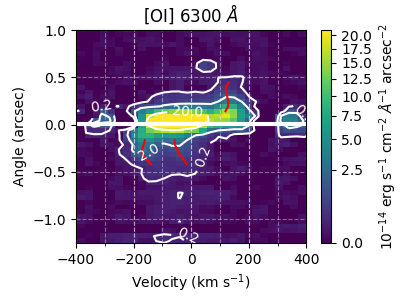}
\includegraphics[width=0.49\textwidth]{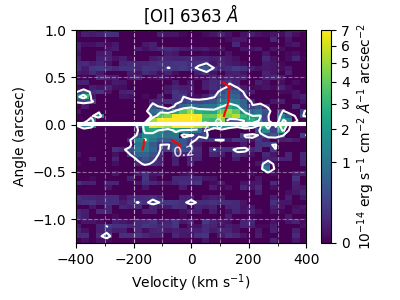}
\caption{Position velocity diagrams (PVD) for DF Tau; the two components of the binary are not resolved and the stellar continuum emission has been subtracted (See text). The x-axis
shows the velocity with respect to the feature rest wavelength in the
velocity frame of the star and the y-axis is the distance along the slit. The
continuum emission from the two stars has been subtracted. The color scale is
chosen to highlight faint emission features; negative values or brighter fluxes
are shown as purple or yellow, respectively. White lines shows the
  position of each star in the binary. Since the two components in DF Tau are unresolved, only one line is visible in this figure. Contours show flux densities in $10^{-14}$~erg~s$^{-1}$~cm$^{-2}$~\AA{}$^{-1}$~arcsec$^{-2}$.
The red lines mark the position of the emission peak for jet components (see text for details).
\label{fig:DFTau}}
\end{center}
\end{figure}
\begin{figure}[h!]
\begin{center}
\includegraphics[width=0.49\textwidth]{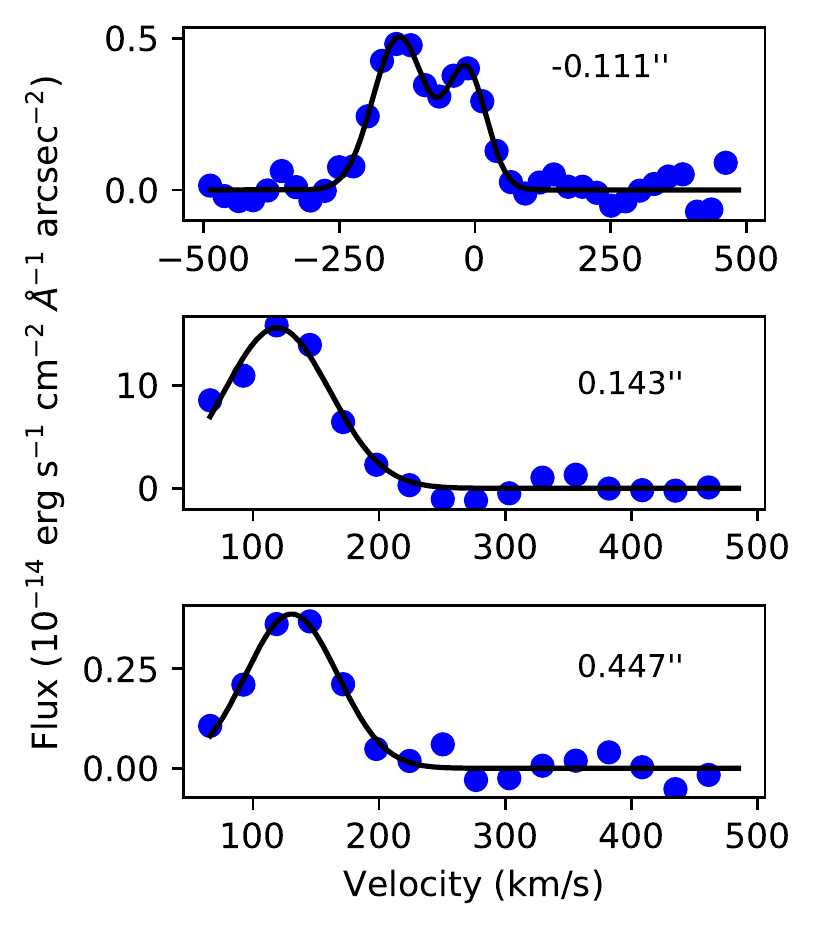}
\caption{Examples of jet component fits for DF Tau for three different
    spatial positions. Data is continuum subtracted close to the 6300~\AA{} line. Statistical errors are smaller than plot symbols, but the true uncertainty on eahc data point is dominated by systematics from the subtraction process. Each plot is labeled with the angular coordinate corresponding to that pixel row. The top plot corresponds to the blue-shifted component with the two Gaussians resolved. The middle plot corresponds to the red-shifted component close to the star, while the bottom plot corresponds to the faintest part of the red-shifted component. The Gaussian fit works well in all of these cases. 
\label{fig:Jet_Fit}
}
\end{center}
\end{figure}

The [O~{\sc i}] position-velocity-diagram (PVD) for DF Tau is shown in figure~\ref{fig:DFTau}
corrected for the stellar radial velocity from \citet{2006AstL...32..759G}.
In order to trace how outflow components accelerate or decelerate with increasing distance from the source, we fit one or more Gaussian emission components to each row of data (each row represents a spectrum at a constant distance from the source). Red lines in the figure mark the
  velocities of the peak of the fitted Gaussians. Examples of these fits are shown in
Fig~\ref{fig:Jet_Fit}. The statistical uncertainty from the fit
is less than the line thickness.
We sum the fluxes and average the $FWHM$ for all fitted Gaussians
further than three pixels from the star center for each outflow component 
(table~\ref{tab:flux}). Uncertainties on the fluxes are dominated by the
subtraction procedure and can be estimated by comparing the fluxes seen in
[O~{\sc i}]~6300\AA{} and [O~{\sc i}]~6363\AA{} lines, whose theoretical ratio
should be 3 \citep{2000MNRAS.312..813S}. Based on the observed ratios, fluxes
are accurate to approximately 50\%.

Both sides of the jet can be traced to a similar distance. The observed
velocity of the high velocity component (HVC) of the approaching jet is about -150 to
-200~km~s$^{-1}$. Since we do not know the inclination angle with respect to
the line-of-sight observed velocities are lower boundaries to the true jet
speed. We also identify a medium velocity component (MVC) around -50~km~s$^{-1}$, but the emission is
weaker than the HVC. There is a red-shifted counter jet at a velocity around
+150~km~s$^{-1}$ and a single emission feature at 50~km~s$^{-1}$, located about
0\farcs6 from the star (72~au projected on the plane of the sky). The HVC of
jet and counterjet is seen to about 0\farcs5 (60~au). All those features are
seen in both [O~{\sc i}] lines, which gives us confidence that even the weaker
features are real detections and not just background fluctuations. We do not
see significant changes in the velocity, thus all outflows must be accelerated
in the inner 0\farcs1 (12~au) where we do not resolve the jet emission.

For a divergent flow with constant velocity, the divergence of the jet
  streamlines is related to the $FWHM$ for the observed FEL. Using the
  simplifying assumption of a homogeneously emitting cone, we can use eqn~5
  from \citet{1990A&A...232...37M} and the values in our
  table~\ref{tab:flux}. 
 The inclination of the DF~Tau outflow cannot be
  directly pole-on, since it is spatially resolved in our data. Assuming an
  inclination $>20$\degr, we derive outflow opening angles $<1$\degr for the
  fast components and $<3$\degr for the slower component. As discussed in
  Section~\ref{sect:intro}, jets from CTTS are typically not homogeneous, but
  this nevertheless shows that we are observing highly collimated jets and not
  wide-angle outflows. Note that the measured values for the $FWHM$ are a
  convolution of the line profile of the FEL and the instrumental resolution
  and thus the true opening angles would be even smaller.

\subsubsection{UY Aur}
\begin{figure}[h!]
\begin{center}
\includegraphics[width=0.49\textwidth]{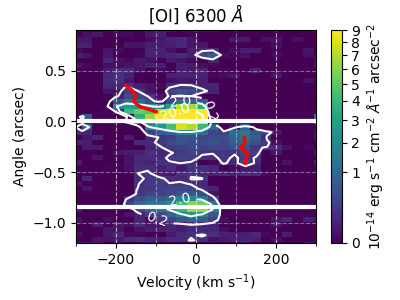}
\includegraphics[width=0.49\textwidth]{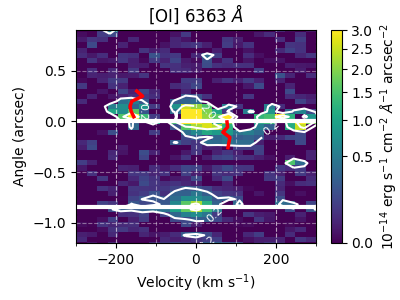}
\caption{Position velocity diagrams for subtracted
image of UY Aur. The stellar continuum emission has been subtracted for both members of the UY Aur binary (see text). The x-axis shows the velocity with respect to the
feature rest wavelength in the velocity frame of the star UY Aur
A and the y-axis the distance along the slit. The position of both components of the UY Aur AB
binary are marked by white lines. The red lines mark the position of the emission peak for jet components (see text for details).
\label{fig:UYAur}
}
\end{center}
\end{figure}

We use the radial velocities from~\citet{2012ApJ...745..119N} for UY~Aur and
show a PVD in figure~\ref{fig:UYAur}.  
Like in DF Tau, the position of
  the peak flux for resolved FEL emission is determined by fitting Gaussians to
  the spectrum at each position and the resulting statistical uncertainties are smaller than
  the width of the red line in the figure.
There is a HVC around -150 to
-200~km~s$^{-1}$ visible to about 0\farcs25 (40~au).This outflow component
seems to originate on the primary and move away from the secondary. There is a
second component, also originating from the primary (the origin on the y-axis
in the figure), that points towards the secondary with a velocity around
+150~km~s$^{-1}$. In [O~{\sc i}]~6366\AA{}, the same components are seen but
the signal is lower. Both plots in the figures also show apparent emission
close to 0~km~s$^{-1}$, but this is an artifact of our subtraction
procedure. The emission is centered on the star. Since our procedure subtracts
the stellar continuum only, the wings of the PSF of the [O~{\sc i}] emission at
the stellar location remain in the images. Fluxes are given in
table~\ref{tab:flux}. 
We
integrate only the resolved signal further than three pixels away from the star
to account for that.
The low $FWHM$ sets strict limits on the opening angle of the
  UY~Aur outflow, showing that we observe highly collimated jets and not
  wide-angle outflows.

\subsection{Upper limits}
The [O~{\sc i}] lines are the only FELs from table~\ref{tab:searchedlines} that fall in the wavelength range of the G750M grating with the available central wavelength settings. We attempt to detect other lines in the G750L data using our subtraction procedure but cannot perform this measurement for the [N~{\sc ii}] lines, because they are too close to the H$\alpha$ line. The [O~{\sc i}] lines are kinematically unresolved, but fluxes are consistent with what is seen in G750M gratings. We inspected figures analogous to figures~\ref{fig:DFTau} and \ref{fig:UYAur} for several regions free of FELs to determine the average noise. We conclude that the [S~{\sc ii}] lines are not detected for any source from table~\ref{tab:obs}. Line-free regions and the [S~{\sc ii}] lines show residuals that decline with distance from the stars with an average of $10^{-14}$~erg~s$^{-1}$~\AA{}$^{-1}$~arcsec$^{-2}$ at 0\farcs2 and 0.05$\times10^{-14}$~erg~s$^{-1}$~\AA{}$^{-1}$~arcsec$^{-2}$ at 0\farcs5. Thus, we conclude that the [O~{\sc i}] lines are at least 3-5 times stronger than [S~{\sc ii}] in DF~Tau and at least two times stronger in UY~Aur. We also checked H$\alpha$ for extension, but did not find extended emission that could indicate a presence of a jet. 

\subsection{Mass loss rates}
With some assumptions on the physical conditions in the jet, we can estimate the total mass loss rate in the resolved outflow components. Following \citet{1994ApJ...436..125H}, we take a plasma with solar abundances emitting close to the peak formation temperature of the [O~{\sc i}] lines. Any jet components that are located outside the
aperture, so hot that oxygen is ionized or so cool that that  [O~{\sc i}] 6300\AA{} is not excited, or with densities above the critical density for [O~{\sc i}] ($10^6$~cm$^{-3}$), are not accounted for in this estimate. Thus, the estimate is a lower limit in the total mass flux in the jet. We use
\begin{eqnarray}
\dot M  & = & 5.95\times10^{-8} \left(\frac{n_e}{10^3\textnormal{ cm}^{-3}}\right)^{-1}\left(\frac{L_{6300}}{10^{-4} L_{\sun}}\right) \nonumber\\
 & & \times \left(\frac{v_{sky}}{100\textnormal{ km s}^{-1}}\right)\left(\frac{l_{sky}}{10^{16}\textnormal{ cm}}\right)^{-1} M_{\sun} \textnormal{ yr}^{-1} \ ,
\end{eqnarray}
which is eqn.~10 from \citet{1994ApJ...436..125H}. 
Since no other FELS are detected, the density can only be constrained to be below the critical density for [O~{\sc i}].
For the following discussion, we will use $n_e =
  10^{3}$~cm$^{-3}$ as the fiducial density because this number was observed in
the jet of DG~Tau \citep{2000A&A...356L..41L}, a well-studied collimated
outflow in the same star forming region, launched from a star of similar age
and only slightly earlier spectral type (K6 vs M0-M2). Several FELs are observed in DG~Tau, thus the density can be calculated from FEL ratios, unlike in our data from DF~Tau and UY~Aur. In practice, the density
likely varies with distance to the source in all jets (as it does in the jet from DG~Tau) and might also
differ between DF~Tau and UY~Aur. All values for the outflow mass flux scale
with $1/n_e$.
Even if the density was known in some
  way (e.g.\ a future observation), the uncertainty in the remaining factors in
  the equation would still make the estimated mass loss uncertain by a factor
  of a few.

\section{Discussion}
\label{sect:discussion}
We detect spatially extended emission in about 10\% of observations. 
The two components of the DF~Tau binary are the strongest accretors in
  the sample \citep{2003ApJ...583..334H}, while UY Aur is close to the
  median. 
If mass flux of the outflow is roughly
  proportional to the accretion rate as suggested by observations \citep{1990ApJ...354..687C,2008ApJ...689.1112C},
  then we would not expect to detect outflows from any source with accretion
  weaker than UY~Aur. There are six binaries in the sample with accretion rates
between UY~Aur and DF~Tau, where a jet could be detected, if it scales with the
accretion rate and happens to be roughly aligned with the slit in the
observations. Assuming that we can detect spatially resolved emission for jets
misaligend with the slit up to about 30~degrees, actually detecting jets in
three out of 16 stars (eight binaries) with mass accretion rates of UY~Aur or
higher, is statistically consistent with all stars having jet emission in a
random orientation.
 However, the fact that the twobinaries where extended
  emission is seen are the same binaries where at least one component shows multiple kinematic
components in the [O~{\sc i}] emission lines \citep{2003ApJ...583..334H}
suggests that the other objects may not have as pronounced outflows.

We compare our detections to other observations of the same jets in the literature to put the properties of these jets into context.

\subsection{DF Tau}

\object{DF Tau} is located at a distance of $125\pm6$~pc
\citep{2016A&A...595A...1G,2018A&A...616A...1G}. It is a binary composed of two
equal mass M2 dwarfs separated by 0.1~arcsec (12.5~au). The primary shows an
infrared excess in its spectral energy distribution (SED), indicating the
presence of a disk, and signatures of accretion, while the secondary seems to
be devoid of circumstellar material \citep{2017ApJ...845..161A}, yet
\citet{2003ApJ...583..334H} determine equal accretion rates of $\dot
M=10^{-7}$~M$_{\sun}$~yr$^{-1}$ for both stars from spectral fitting, which is
about an order of magnitude above the mass loss rate of all outflow components
we detect here assuming a density in the outflow of $n_e=10^3$~cm$^{-3}$.
\citet{2004ApJ...609..261H} observed the jet of DF Tau with low-resolution slitless spectroscopy with STIS. They detect a jet and counterjet at a
position angle of 127~degrees, but the binary is not resolved and it was unclear at the time which star was the origin of which outflow component. Given the absence of circumstellar material around the secondary component, it seems very likely now that we are looking at a bipolar jet launched from the primary star. 

In the observations presented here, the position angle of the slit was 153~degrees, only 26~degrees from the jet axis. This means that for distances beyond about 0\farcs5, a well-collimated outflow would not be visible any longer in our images because it reaches the edge of the 0\farcs2 wide slit and indeed little signal is detected at larger distances from the star. The LVC seen at 0\farcs6 must therefore be part of a larger, possibly less collimated, structure. When the emission is not centered in the slit, the wavelength scale is not accurate. However, a slit half-width of 0\farcs1 corresponds to only 2 pixels (about 50~km~s$^{-1}$). Given the size of the PSF, even a source located on the edge of the slit would be detected in several pixels, so the maximal velocity shift that can be explained by the emission not being centered in the slit is about $\pm30$~km~s${-1}$.

This data was taken about one and two years before the two slitless exposures of \citet{2004ApJ...609..261H} respectively. We trace the jet out to slightly larger distances and resolve a HVC and MVC on both sides of the jet, while the slitless data does not allow a velocity measurement.

\subsection{UY Aur}

\object{UY Aur} is located at a distance of $156\pm2$~pc
\citep{2016A&A...595A...1G,2018A&A...616A...1G}. It again is a binary system
with two components of similar mass \citep[M0 and M2,
  see][]{2003ApJ...583..334H} but a much larger separation (projected 140~au)
than DF Tau. \citet{2003ApJ...583..334H} determine very similar mass accretion
rates around $\dot M=2\times10^{-8}$~M$_{\sun}$~yr$^{-1}$, which, like in
DF~Tau, is about one to two orders of magnitude higher than the summed mass
outflow rates we measure assuming $n_e=10^3$~cm$^{-3}$.  UY~Aur~A has a disk
resolved by ALMA with a position angle of $177\pm11$~degrees and an inclination
of $56\pm16$ degrees \citep{2014ApJ...784...62A}. FELs in a jet were observed
in 1988 by \citet{1997A&AS..126..437H} in ground-based observations with a
position angle around 40~degrees. They trace the emission out to several arcsec
in [O~{\sc i}]~6300~\AA{}. The spectral and spatial resolution of their data is
not as good as what we present here, however, they trace the jet out to larger
distances. Their data indicate the presence of different emission components
ranging from -240~km~s$^{-1}$ in the approaching jet to +180~km~s$^{-1}$ in the
receding jet. More recently \citet{2014ApJ...786...63P} and
\citet{2019ApJ...884..159B} performed adaptive-optics observations with an
integral field unit (IFU) in the IR and they obtain detailed images in the
[Fe~{\sc ii}] $\lambda$1.257~$\mu$m and $H_2$ $\lambda$2.12~$\mu$m
  lines. They detect blue-shifted emission in a circular region around the
primary as well as in a ``bridge'' that connects the primary and the secondary
(offset by about 0\farcs2 to the north from the direct line connecting the
primary and the secondary); redshifted emission is seen close to the primary on
the side that faces towards the secondary as well as in the ``bridge''
region. Together with the absorption features observed in the stellar spectra,
this suggests a geometry where the primary has a wide-angle, fast wind on both
sides and a more collimated, red-shifted jet on the far side of the disk, in
the direction of the secondary. The secondary does not have a wide-angle wind
in [Fe~{\sc ii}] but only a collimated jet. In the ``bridge'' region we see the
red-shifted jet of the primary and the blue-shifted jet of the secondary
projected onto the same area of the sky.

The long-slit in our data contains both stars of the binary and overlaps the ``bridge'' region in between. The slit is 0\farcs2 wide, so it contains some of the flux in the ``bridge'', but not the peak of the flux distribution. Looking just at the region covered by our slit, we see remarkably similar emission in [O~{\sc i}] compared with the [Fe~{\sc ii}] observations of \citet{2014ApJ...786...63P}. We also have indications for a fast, blue-shifted outflow ``above'' UY Aur~A and we see both red-and blue-shfited emission in the ``bridge'' region, again with velocities similar to those seen in [Fe~{\sc ii}]. Given that the data from \citet{1997A&AS..126..437H} was taken in 1988, this data in 1998, and the \citet{2014ApJ...786...63P} data in 2007 and that the feature with a velocity of 200~km~s$^{-1}$ would move about 3\arcsec{} on the sky in ten years, this indicates all features seen represent stable outflow patterns. This includes the blue-shifted outflow from UY~Aur~A, interpreted as a fast, wide angle wind by \citet{2014ApJ...786...63P}, the red-shifted outflow from UY Aur~A (interpreted as a fast, wide-angle wind, too), and the blue-shifted outflow from UY~Aur~B (identified as a jet).

\subsection{Comparison to other jets}
\citet{2008ApJ...689.1112C} presented HST/STIS observations of five jets from
T~Tauri stars, covering a very similar wavelength range as the data we analyze
here. While they see morphological features that our data would not resolve,
all those jets are overall similar to those we present in this paper. In
particular, they also find outflow velocities up to 200~km~s$^{-1}$ in [O~{\sc
    i}] and can resolve the [O~{\sc i}] emission lines into a fast and a slow
component in some cases. Because they observe FELs from more than one species,
they can estimate the density and find typical values around
$n_e=10^4$~cm$^{-3}$. With that number, they can estimate the ratios of
accretion and mass loss rate in their objects to 0.01-0.07. Assuming that the
densities in the outflows from DF~Tau and UY~Aur are similar, the ratio of
accretion to outflow rates that we estimated above fall into the same range. 

As already mentioned in Sect.~\ref{sect:intro}, [O~{\sc i}] is not the only
species where we observe outflows that are clearly split into different
velocity components, e.g.\ C~{\sc iv} which traces a hotter outflow component
is often seen around 200~km~s$^{-1}$, for example in DG~Tau
\citep{2013A&A...550L...1S} and RY~Aur \citep{2018ApJ...855..143S}. This
component is typically seen only within a few tens of au, and it roughly matches the
velocity of the HVC we detect in [O~{\sc i}]. We can thus conclude, that the
jets of DF~Tau and UY~Aur that we present here is fairly typical for T~Tauri stars.

Only a small number of binary CTTS are known to drive jets. This
  includes the eponymous \object{T Tau} itself, a hierarchical multiple system
  where the main components are separated by $>100$~au, drive separate outflows \citep{1999ApJ...523..709S}.
 \object{RW Aur} is also resolved with a separation $>100$~au  and only RW~Aur~A drives
 a visible jet \citep{2018ApJ...855..143S}. In contrast, \object{KH 158D}
 (separation $<20R_*$), does not have individual jets
 \citep{2010ApJ...708L...5M}. We confirm jet emission from DF~Tau (separation
 of order 10~au) and UY~Aur (separation of order 100~au). DF~Tau seems
 like a good candidate to test the predictions from \citet{2018ApJ...861...11S}
 about the gravitational influence of one star on the outflow of the other, but
 our data does not resolve the shape of the jet spatially, so that it remains
 unclear if the observed velocity variations in the outflow
 (Figure~\ref{fig:DFTau}) are intrinsic to the launching or the result of
 projection effects in a precessing jet. The outflows in UY~Aur are more
 complex \citep{2014ApJ...786...63P} with both narrow and wide-angle components in different alignments and
thus seem more similar to the random orientation that simulations predict for
the alignment in wide binaries \citep{2016ApJ...827L..11O}.

\section{Summary}
\label{Sect:summary}
We present a search for resolved emission in FELs in a sample of twenty one long-slit HST/STIS observations. We detected resolved [O~{\sc i}] emission in the binaries of DF~Tau and UY~Aur. In DF~Tau, we see a HVC and a MVC in both the jet and the counterjet. Both are detected as far out as possible, given that our slit position angle is misaligned compared to the jet direction. The HVC are accelerated within about 15~au and retain a constant velocity after that. In UY~Aur, a complex outflow geometry with components originating on both members of the binary was known before. We detect all components that overlap with the spatial position of our slit, indicating that the outflow pattern is stable for at least one decade.

\acknowledgments
Support for this work was provided for AVU and HMG by NASA through
grant GO-13766.010 from the Space Telescope Science Institute. This research made use of Astropy,\footnote{http://www.astropy.org} a community-developed core Python package for Astronomy \citep{2013A&A...558A..33A,2018AJ....156..123A}. 
This work has made use of data from the European Space Agency (ESA) mission
{\it Gaia} (\url{https://www.cosmos.esa.int/gaia}), processed by the {\it Gaia}
Data Processing and Analysis Consortium (DPAC,
\url{https://www.cosmos.esa.int/web/gaia/dpac/consortium}). Funding for the DPAC
has been provided by national institutions, in particular the institutions
participating in the {\it Gaia} Multilateral Agreement.

\facilities{HST (STIS), Gaia}
\software{Astropy \citep{2013A&A...558A..33A,2018AJ....156..123A}}

\bibliographystyle{aasjournal}
\bibliography{bib}

\end{document}

%% file: table_of_observations.tex
\begin{table*}
\caption{Log of observations\label{tab:obs}}
\begin{center}
\begin{tabular}{cccccc}
\hline\hline
OBSID G750L & OBSID G750M & Target & Observation Date & Aperture & Position angle (deg) \\
\hline
o55101010& o55101020 & FO TAU & 1999-10-17 & 52X0.2 & 50.15 \\

%o55101020 & FOTAU & 1999-10-17 & 52X0.2 & 50.15 \\

o55102010& o55102020 & DD TAU & 1999-09-05 & 52X0.5 & 64.54 \\

%o55102020 & DDTAU & 1999-09-05 & 52X0.5 & 64.54 \\

o55104010& o55104020 & UZ TAU-W & 1999-11-02 & 52X0.2 & 52.54 \\

%o55104020 & UZTAU-W & 1999-11-02 & 52X0.2 & 52.54 \\

o55105010& o55105020 & LKCA 7 & 1999-01-25 & 52X0.2 & 250.0 \\

%o55105020 & LKCA7 & 1999-01-25 & 52X0.2 & 250.0 \\

o55106010& o55106020 & LKHA 332 & 1998-12-23 & 52X0.2 & 251.0 \\

%o55106020 & LKHA332 & 1998-12-23 & 52X0.2 & 251.0 \\

o55107010& o55107020 & UY AUR & 1998-12-25 & 52X0.2 & 271.0 \\

%o55107020 & UYAUR & 1998-12-25 & 52X0.2 & 271.0 \\

o55108010& o55108020 & 040047+2603W & 1998-12-04 & 52X0.2 & 272.0 \\

%o55108020 & 040047+2603W & 1998-12-04 & 52X0.2 & 272.0 \\

o55109010& o55109020 & FQ TAU & 1998-12-05 & 52X0.5 & 303.0 \\

%o55109020 & FQTAU & 1998-12-05 & 52X0.5 & 303.0 \\

o55110010& o55110020  & FS TAU & 2000-12-13 & 52X0.2 & 309.0 \\

%o55110020 & FSTAU & 2000-12-13 & 52X0.2 & 309.0 \\

o55111010& o55111020 & HARO6-28 & 1998-12-02 & 52X0.2 & 290.1 \\

%o55111020 & HARO6-28 & 1998-12-02 & 52X0.2 & 290.1 \\

o55112010& o55112020 & LKHA332G2 & 1998-12-12 & 52X0.5 & 269.4 \\

%o55112020 & LKHA332G2 & 1998-12-12 & 52X0.5 & 269.4 \\

o55113010& o55113020 & 040142+2150 & 1998-12-01 & 52X0.2 & 289.0 \\

%o55113020 & 040142+2150 & 1998-12-01 & 52X0.2 & 289.0 \\

o55114010& o55114020 & IS TAU & 2000-12-03 & 52X0.2 & 317.0 \\

%o55114020 & ISTAU & 2000-12-03 & 52X0.2 & 317.0 \\

o55115010& o55115020 & FV TAU & 2000-12-03 & 52X0.5 & 322.4 \\

%o55115020 & FVTAU & 2000-12-03 & 52X0.5 & 322.4 \\

o55116010& o55116020 & FV TAU-C & 1998-12-02 & 52X0.2 & 337.0 \\

%o55116020 & FVTAU-C & 1998-12-02 & 52X0.2 & 337.0 \\

o55117010& o55117020 & GH TAU & 2000-12-01 & 52X0.2 & 350.8 \\

%o55117020 & GHTAU & 2000-12-01 & 52X0.2 & 350.8 \\

o55118010& o55118020 & LKHA 331 & 1999-10-08 & 36X0.6P45 & 100.0 \\

%o55118020 & LKHA331 & 1999-10-08 & 36X0.6P45 & 100.0 \\

o55119010 & o55119020 & UZ TAU-W-OFF & 2000-09-04 & 52X0.5 & 64.79 \\

%o55119020 & UZTAU-W-OFF & 2000-09-04 & 52X0.5 & 64.79 \\

o55120010& o55120020 & DF TAU & 1998-11-26 & 52X0.2 & 18.04 \\

%o55120020 & DFTAU & 1998-11-26 & 52X0.2 & 18.04 \\

o55121010& o55121020 & XZ TAU & 2000-12-02 & 52X0.2 & 194.1 \\

%o55121020 & XZTAU & 2000-12-02 & 52X0.2 & 194.1 \\

o55122010& o55122020 & V807 TAU & 1998-11-30 & 52X0.2 & 19.03 \\

%o55122020 & V807TAU & 1998-11-30 & 52X0.2 & 19.03 \\
\hline
\end{tabular}
\end{center}
\end{table*}